\documentclass[11pt]{article}
\usepackage[latin1]{inputenc}
\usepackage[T1]{fontenc}
\usepackage{amsmath}
\usepackage{amsfonts}
\usepackage[]{epsfig}
\usepackage{amssymb}
\usepackage{color}
\usepackage{graphicx}
\usepackage{tikz}
\usepackage{hyperref}

  \newcommand{\be}{\begin{equation}}
\newcommand{\ee}{\end{equation}}

\newcommand{\ba}{\begin{eqnarray}}
\newcommand{\ea}{\end{eqnarray}}
\newcommand{\oh}{\displaystyle{\frac{1}{2}}}
 
\newcommand{\bea}{\begin{eqnarray}}
\newcommand{\eea}{\end{eqnarray}}
\newcommand{\id}{\!\!\not\!\partial}

\newcommand{\fs}{i\kern+.01em\hbox{\raise.20ex\hbox{$/$}\kern-.58em$s$}}

\newcommand{\bs}{i\kern-.01em\hbox{\raise.25ex\hbox{$/$}\kern-.52em$b$}}
\newcommand{\qs}{/\kern-.52em s}

\newcommand{\dd}{\kern.06em\hbox{\raise.25ex\hbox{$/$}\kern-.40em$\partial$}}

\begin{document}
\tikzstyle{bag} = [text width=2em, text centered]
\tikzstyle{bag1} = [text width=5em, text centered]
\tikzstyle{end} = []
\title{Induced parity-odd effective action for a Dirac field on ${\mathcal
S}^2 \times {\mathcal S}^1$}
\author{C.~D.~Fosco$^a$   and
F.~A.~Schaposnik$^b$\thanks{Associated with CICBA}
\\
~
\\
~
\\
{\normalsize $^a\!$\it Centro At\'omico Bariloche and Instituto Balseiro,}\\
{\normalsize $\!$\it Comisi\'oon Nacional de Energ\'\i a At\'oomica, 8400 Bariloche, Argentina}\\
{\normalsize $\!$\it }\\
{\normalsize $^b\!$\it Departamento de F\'\i sica, Universidad
Nacional de La Plata}\\ {\normalsize\it Instituto de F\'\i sica La Plata-CONICET}\\
{\normalsize\it C.C. 67, 1900 La Plata,
Argentina}
 }

\date{\today}

\maketitle
%===================================================================
\begin{abstract}
We evaluate the parity-odd part of the effective action due to massive
Dirac fermions on a ${\mathcal S}^2 \times {\mathcal S}^1$ manifold,
minimally coupled to an external Abelian gauge field. We do that for a
special class of gauge-field configurations, which is specially suitable
to the study of the behaviour of the fermionic determinant under large
gauge field configurations, which are allowed by the space-time geometry.
\end{abstract}
%=======================================================
\section{Introduction}
Physical systems involving topological quantum matter (see \cite{Hasan} and
reference therein) have recently aroused much attention and attracted
intense theoretical research, in part because of the existence of
dualities in their effective field theory model descriptions. The latter
usually involve Chern-Simons (CS) matter theories coupled to scalars, and
similar theories coupled to fermions. Originally discussed in \cite{Widual}-\cite{Nastase} these field theoretical investigations were extended in several directions \cite{directions}. In those
models, two out of the three spatial dimensions of the real material are
compactified to the ${\mathcal S}^2$ sphere, which is meant to describe a
boundary (the third spatial coordinate is assumed to have
a finite extension).  Finally, in order to account for finite-temperature
effects, the remaining, Euclidean time coordinate is an ${\mathcal S}^1$
circle of circumference $\beta = T^{-1}$, $T$ denoting the temperature (we
adopt units such that Boltzmann's constant $k_B \equiv 1$).

In this way, phenomena which take place on the boundary of topological
insulators or superconductors can be studied using models with bosons or
fermions, coupled to external and dynamical gauge fields,  with a CS action
for the former and BF terms for their interaction.

Based on the previous motivation, systems described by actions defined on a
manifold ${\mathcal M} = {\mathcal S}^2 \times {\mathcal S}^1$, involving
bosonic or fermionic matter minimally coupled to an external gauge field,
have been studied.  This has lead us to consider, in this paper, a
particularly interesting object, namely, the effective action $\Gamma[A]$
for a massive Dirac field on precisely that kind of spacetime manifold.

It is a well-known property that $\Gamma[A]$, for Dirac fermions in
\mbox{$d=3$} (either Euclidean or Minkowskian) spacetime dimensions,
contains a parity and time-reversal breaking term $\Gamma_{\rm odd}[A]$.
This is a reflection of the breaking of those symmetries; explicitly, when
there is a finite fermion mass $m$ at the classical level, or implicitly by
quantum effects, due to the (unavoidable) regularization.  The main
difference between those two contributions, besides their different
origins, is that the one due to the regularization is local and independent
of $m$, while the one
corresponding to the explicit introduction of a mass term is local, and
proportional to the CS action, only when $m$ tends to
infinity~\footnote{Or, equivalently, when keeping the leading terms in an
expansion in derivatives.}.  The structure of the mass-dependent,
parity-odd term in the effective action at finite temperature cannot be
determined just by symmetry considerations, but requires an explicit
computation \footnote{Let us note that under certain special conditions,
	out of our scope, a regularization involving the  $\eta[A]$ invariant should be
	consider \cite{Widual}.}.
The anomalous and the mass-dependent contributions to $\Gamma_{\rm odd}[A]$ will be termed
$\Gamma_{\rm odd}^{(0)}[A]$ and $\Gamma_{\rm odd}[A,m]$, respectively, so that
\begin{equation}
\Gamma_{\rm odd}[A] \;=\; \Gamma_{\rm odd}^{(0)}[A]\,+\,\Gamma_{\rm odd}[A,m] \;.
\end{equation}

The regularization procedure produces a global sign ambiguity in
$\Gamma_{\rm odd}^{(0)}[A]$. Indeed, within the $\zeta$-function
regularization approach, this ambiguity is associated to the choice of a
contour when defining the complex powers of the Dirac operator on
odd-dimensional manifolds~\cite{GRS} and is independent of $m$. Using this
approach for fermions coupled to a $U(1)$ gauge field background in
$3$-dimensional Euclidean space, with the effective action defined by
\begin{equation}
\exp(-\Gamma[A]) \equiv \int D\bar\psi D\psi \exp\left(
\int_{{\mathbb R}^3}\bar\psi (~\id  +i\not\!\!A + m)\psi
\right)
\label{dospartecero}
\end{equation}
the result for the anomalous part has been shown to be \cite{GRS}:
\be
\Gamma_{\rm odd}^{(0)}[A] = \pm  \frac{i}{2} \, S_{CS}[A]
\label{odd0}
\ee
where
\be
S_{CS} = \frac1{4\pi}\int d^3x
\varepsilon^{\mu\nu\alpha}A_\mu \partial_\nu A_{\alpha} \;.
\label{CheCS}
\ee

Regarding the mass-dependent part of the parity-odd effective action
$\Gamma_{\rm odd}[A,m]$ one consistently obtains, either by
using the $\zeta$-function approach or a derivative
expansion for the external gauge field, the result:
\be
\Gamma_{\rm odd}[A,m] = \frac{i}{2} \,\frac{m}{|m |} S_{CS}[A] \;+\;
{\mathcal O}(m^{-2}) \;.
\label{oddm}
\ee

Note that, by an appropriate choice of signs,  one can cancel  the $1/2$ factor in
$\Gamma_{\rm odd}[A]$ leading to an odd-parity effective action which is
gauge invariant even for large gauge transformations. It is important to
realize that the leading term in a mass expansion is the most relevant one
when considering topological properties, like invariance under those large
transformations.

In other regularization schemes, the same phenomenon must and does manifest
itself; for instance, in the Pauli-Villars approach, the breaking may be traced back
to the presence of a regulator, a spinorial field with a Dirac-like
action, but quantized using Bose-Einstein statistics.  The sign of the
mass of the regulator is not determined by the regularization.  Removing
the regularization at the end of the calculations amounts, in this case,
to letting the regulator mass tend to infinity. This process yields a
finite limit, the CS action, but with an overall sign which depends on the
sign of the regulator mass.

It is the aim of this paper to evaluate the mass-dependent term
$\Gamma_{\rm odd}[A,m]$ for massive Dirac fermions on the manifold
${\mathcal M} = {\mathcal S}^2 \times {\mathcal S}^1$. The anomalous part
does indeed exist, and it is part of $\Gamma_{odd}[A]$, but is has a
well-known form: it is the integral of the CS $3$-form on ${\mathcal M}$,
with the same coefficient as in the $T=0$ case.

This paper is organized as follows: in Section~\ref{sec:review}, we briefly
review known results for the effective action corresponding to a massive
Dirac fermion on the ${\mathbb R}^2 \times {\mathcal S}^1$ manifold. Based
on this approach, in Sect.~\ref{sec:sphere} we consider the case of
${\mathcal S}^2 \times {\mathcal S}^1$. In Section~\ref{sec:disc} we
present a discussion of our results.
%%%%%%%%%%%%%%%%%%%%%%%%%%%%%%%%%%%%%%%%%%%%%%%%%%%%%%%%%%%%%%%%%%%%%%%%%%%%%%%
%%%%%%%%%%%%%%%%%%%%%%%%%%%%%%%%%%%%%%%%%%%%%%%%%%%%%%%%%%%%%%%%%%%%%%%%%%%%%%%
%%%%%%%%%%%%%%%%%%%%%%%%%%%%%%%%% Review %%%%%%%%%%%%%%%%%%%%%%%%%%%%%%%%%%%%%%
%%%%%%%%%%%%%%%%%%%%%%%%%%%%%%%%%%%%%%%%%%%%%%%%%%%%%%%%%%%%%%%%%%%%%%%%%%%%%%%
%%%%%%%%%%%%%%%%%%%%%%%%%%%%%%%%%%%%%%%%%%%%%%%%%%%%%%%%%%%%%%%%%%%%%%%%%%%%%%%
\section{Massive Dirac fermions at finite temperature}
\label{sec:review}
The study the effective action for a Dirac field in $2+1$ dimensions at
finite temperature, in the Matsubara formalism one should regard the field
as living on Euclidean ${\mathbb R}^2\times {\mathcal S}^1$ space. In this
case, gauge invariance under large gauge transformations associated to the
${\mathcal S}^1$  ``time'' is spoiled at any finite order in a perturbative
calculation in powers of the gauge coupling constant. That invariance under
large gauge transformations can, however, be rescued, if one follows a
non-perturbative approach, like the ones presented in~\cite{A}-\cite{C}.

The key idea in refs.~\cite{B}-\cite{C} was to first write the gauge invariant
parity-odd part  $\Gamma_{\rm odd}$ of the effective action $\Gamma$ due to
the fermions in the presence of an external gauge field, to then reduce
the problem to the calculation of a set of Fujikawa Jacobians in $R^2$
space and finally to obtain $\Gamma_{\rm odd}$ by summing over
Matsubara modes associated to the discrete Fourier transformations in the
``time variable''.

The above ``reduction'' mechanism  whereby the problem was essentially
reduced to a collection of $1+1$ systems was implemented for a certain class
of gauge field configurations,
\be\label{eq:gauge}
A_i = A_i(x) \;,\;\;\;\; A_3 \,=\, A_3(\tau) \;,
\ee
where we have adopted the convention (to be followed in the rest of this
paper) to denote by  $x$ just the two spatial coordinates, namely $x =
(x_1,x_2)$. We will also assume that indices from the middle of the Latin alphabet
will run from $1$ to $2$, while if they belong to the Greek one, their range
is from $1$ to $3$. The third Euclidean coordinate will be alternatively
denoted as $\tau$ or $x_3$.

Note that (\ref{eq:gauge}) may be regarded as a gauge-fixed version of the
gauge-invariant conditions
\begin{equation}
	F_{3i} \;=\; 0 \;,\;\;\; F_{ij} \;=\; F_{ij}(x) \;,	
\end{equation}
which correspond to field configurations which, if the gauge field were
determined by a Maxwell Lagrangian in $2+1$ dimensions, would be those of
pure magnetostatics. Namely, configurations produced by a vanishing charge
density and a solenoidal time-independent current $J_i(x)$ which determines
$F_{ij}$ via Amp\`ere's Law:
\begin{equation}\label{eq:amp}
	\partial_j F_{ji}(x) \;=\; - J_i(x) \;.
\end{equation}
Since $\partial_i J_i = 0$, we may write it as the curl of a pseudoscalar
$\zeta(x)$, namely:  \mbox{$J_i(x) = \epsilon_{ij} \partial_j \zeta(x)$}.  In terms of the
magnetic field $B(x) = \oh \epsilon_{ij} F_{ij}(x)$, we see that (\ref{eq:amp})
becomes:
\begin{equation}
	B(x) \;=\; F_{12}(x) \;=\;  \zeta(x) \;.
\end{equation}

Let us now describe the finite temperature calculation. In the
path-integral framework,  one works with an Euclidean time in the
interval $[0,\beta]$, imposing anti-periodic boundary conditions to
the fermions in that interval.  Hence the manifold becomes
$\mathbb{R}^2\times {\mathcal S}^1$ and then one sees that invariance under
large gauge transformations associated to the ${\mathcal S}^1$  domain
is spoiled at any finite order in the gauge coupling constant (see
\cite{Dunne} and references therein). This fact has caused some confusion
regarding the CS coefficient discreteness condition, since the coefficient of
the CS action would then satisfy a rather unphysical $\beta$-dependent
quantization condition in conflict with large gauge invariance of the effective action.

The problem has been solved by following non-perturbative approaches. We shall
briefly describe here the procedure followed in refs.~\cite{B}-\cite{C} to
perform such calculation. The main idea was to  reduce the problem to the
calculation of a set of Fujikawa Jacobians on an ${\mathbb R}^2$ space, and then
to obtain the complete partition function by summing over Matsubara modes
associated to the Fourier transformations on ${\mathcal S}^1$.

The mass-dependent~\footnote{Note that only the mass-dependent part of
$\Gamma_{\rm odd}$ may depend non-trivially on $\beta$.} part of
$\Gamma_{\rm odd}$ may be obtained by using the expression:
\be
\Gamma_{\rm odd}[A,m] = \frac12 \left( \Gamma[A,m] - \Gamma[A,-m]\right)
\;,
\label{unaparte}
\ee
for a system defined by the action
\be
S = \int_0^\beta d\tau\int_{\mathbb{R}^2}\bar\psi (~\id  +i\not\!\!A +
m)\psi \;.
\ee
The result has been shown to be:
\bea
\Gamma_{\rm odd} [A,m] &=& \frac{i}{2\pi} \frac{m}{|m|}
\arctan\left[ \tanh \left(\frac{\beta |m|}2\right)
\times
\tan\left(\frac1{2}\int_0^\beta d\tau A_3(\tau) \right)  \right]\nonumber\\
&&\times \int_{{\mathbb R}^2} d^2xF_{12} \;.
\label{resul}
\eea
This result, obtained for the class of configurations explicited in
(\ref{eq:gauge}), has the correct zero-temperature limit:
\be
\lim_{\beta \to \infty} \Gamma_{\rm odd}^{I}[A] =  \frac{i}2 \frac{m}{|m |}
\frac1{4\pi}\int d^3x \varepsilon^{\mu\nu\alpha}A_\mu \partial_\nu
A_{\alpha} = \frac{i}2\frac{m}{|m |} S_{CS} \;.
\label{clear}
\ee
The result in Eq.\eqref{clear} is not invariant by itself under
large gauge transformations, because of the $1/2$ factor affecting the CS term. The same
happens in the finite temperature result as given by eq.\eqref{resul}.
Indeed, if one makes a large gauge transformation  $\Omega_n(x_1,x_2,\tau)$
with $n = 2p +1$ that winds an arbitrary number $n$ of times around cyclic
time direction,
\be
\Omega_n(\beta,x) = \Omega_n(0,x) + 2\pi n \;, \;\;\; n \in {\mathbb Z}
\label{catorce}
\ee
the argument  in the tangent of eq. \eqref{resul} is shifted by
$(2p+1)\pi$. Although the tangent is not sensitive to such a change, one
has to keep track of it by shifting the branch used for the $\arctan$
definition so that  $\Gamma_{\rm odd}[A]$ changes by the addition of
$in\pi$ under a large gauge transformation and for odd  $n = 2k+1$  one should
correct this with an appropriate gauge invariant regularization.

Note that this gauge non-invariance can be compensated by a judicious
choice of the sign in the anomalous term, which has an analogous
behaviour under the same transformations.

Some properties of $\Gamma_{\rm odd}[A,m]$  may also be studies by
considering its contribution $\theta_{\rm odd}[A,m]$ to the phase of the fermionic determinant,
\begin{equation}
	e^{i \theta_{\rm odd}[A,m]} \;\equiv \;	e^{-\Gamma_{\rm odd}[A,m]}
\end{equation}
where the angle $\theta_{\rm odd}[A,m]$ can be written as follows:
\begin{equation}
\theta_{\rm odd}[A,m] \,=\, n_B \; {\rm arg}(z)
\end{equation}
where
\begin{equation}
	z \;\equiv\; \cos \big(\frac{1}{2} \int_0^\beta d\tau [ A_3(\tau) +
	i m ] \big) \;,
\end{equation}
and $n_B \;\equiv\; \frac{1}{2\pi} \int_{{\mathbb R}^2}d^2xF_{12}$, which is
quantized for non-singular configurations, namely, $n_B \in {\mathbb Z}$.

We conclude this review Section commenting the fact that the previous
result for $\Gamma_{odd}$ may be generalized to a slightly more
general class of configurations. Indeed, having in mind the fact that the
ones considered before were purely magnetostatic ones, we can think about
the generalization to the case of having static magnetic plus electric
fields:
\begin{equation}
	F_{3i} \;=\; F_{3i}(x) \;,\;\;\;\; F_{ij} \;=\; F_{ij}(x) \;.	
\end{equation}
In this case, the only change one has to introduce in the derivation of the
previous case is the fact that the anomalous Jacobian corresponds to a
space-dependent phase, and the result becomes the straightforward generalization:
\be \label{resul1}
\Gamma_{\rm odd} [A,m] = i \frac{m}{|m|} \int_{{\mathbb
R}^2}d^2x {\cal L}[A,m,\beta]
\ee
where
\be
{\cal L}[A,m,\beta] =\frac1{2\pi} \arctan\left[ \tanh \left(\frac{\beta |m|}2\right)
\tan\left(\frac1{2}\int_0^\beta d\tau A_3(x,\tau) \right)  \right]
F_{12}(x) \;.
\ee

%%%%%%%%%%%%%%%%%%%%%%%%%%%%%%%%%%%%%%%%%%%%%%%%%%%%%%%%%%%%%%%%%%%%%%%%%%%%%%%
%%%%%%%%%%%%%%%%%%%%%%%%%%%%%%%%%%%%%%%%%%%%%%%%%%%%%%%%%%%%%%%%%%%%%%%%%%%%%%%
%%%%%%%%%%%%%%%%%%%%%%%%%%%%%%%%% sphere %%%%%%%%%%%%%%%%%%%%%%%%%%%%%%%%%%%%%%
%%%%%%%%%%%%%%%%%%%%%%%%%%%%%%%%%%%%%%%%%%%%%%%%%%%%%%%%%%%%%%%%%%%%%%%%%%%%%%%
%%%%%%%%%%%%%%%%%%%%%%%%%%%%%%%%%%%%%%%%%%%%%%%%%%%%%%%%%%%%%%%%%%%%%%%%%%%%%%%
\section{Massive Dirac field on ${\mathcal S}^2 \times {\mathcal S}^1$:
the determinant}\label{sec:sphere}
We consider now a massive Dirac field $\psi,\,{\bar\psi}$, living on
the  manifold\\ \mbox{${\mathcal M} \equiv S^2 \times S^1$}, where
$S^1$ is a periodic (imaginary) time coordinate, while $S^2$
denotes a spatial sphere of radius $R$.

The manifold ${\mathcal M}$ is equipped with the product metric, namely
\begin{equation}\label{eq:met}
	ds^2 \;=\; (d\tau)^2 + g_{ij}(x) dx^i dx^j \;,
\end{equation}
which corresponds to a particular case of static space-time.

For a field defined on ${\mathcal M}$, in the presence of
an external Abelian gauge field $A$, we are interested in evaluating the
imaginary part of the Euclidean effective action, $\Gamma(A,m)$, where:
\begin{equation}\label{eq:defgamma}
	e^{- \Gamma(A,m)} \;=\; \int {\mathcal D}\psi {\mathcal D}\bar\psi
	\; e^{- {\mathcal S}(\bar\psi,\psi,A,m)} \;,
\end{equation}
where the Dirac field is again assumed to be minimally coupled to an Abelian gauge
field $A_\mu(x)$, so that the action ${\mathcal S}$ is defined as follows:
\begin{equation}\label{eq:defs}
	{\mathcal S}(\bar\psi,\psi,A,m)\;=\; \int_{\mathcal M} d^3x
	\,\sqrt{g} \; {\bar\psi}(x) \big( {\mathcal D}_{\mathcal M}\,+\,m \big) \psi(x) \;,
\end{equation}
where ${\mathcal D}_{\mathcal M}$ denotes the Dirac operator corresponding
to ${\mathcal M}$ and coupling to the gauge field $A_\mu$.

Some objects appearing in the action involve the geometry of ${\mathcal M}$
in a more specific way: $g \equiv \det(g_{\mu\nu}) = \det(g_{ij})$, where
$g_{\mu\nu}$ may be represented in matrix form:
\begin{equation}
	\big( g_{\mu\nu} \big) \;=\;
	\left(
		\begin{array}{ccc}
			g_{11}({\mathbf x}) & g_{12}({\mathbf x}) & 0 \\
			g_{21}({\mathbf x}) & g_{22}({\mathbf x}) & 0 \\
			0 & 0 & 1	
		\end{array}
	\right) \;.
\end{equation}
Denoting by $x^i \to {\mathbf r}(x^1,x^2)$  a parametrization of the surface
${\mathcal S}^2$:
\begin{equation}
{\mathbf r}(x^1,x^2) \in {\mathbb R}^3 \;,\;\;
{\mathbf r}(x^1,x^2) \cdot{\mathbf r}(x^1,x^2) = R^2
\end{equation}
(where the dot denotes the Euclidean scalar product in ${\mathbb R}^3$), we
have the spatial components $g_{ij}$ of the induced metric tensor:
\begin{equation}
	g_{ij} \;=\; \partial_i{\mathbf r} \cdot \partial_j{\mathbf r} \;.
\end{equation}
Besides, note that, because of the special form of the metric tensor above, the only
non-trivial vierbeins are the two corresponding to $S^2$ (they are moreover
time-independent).

Following the strategy of refs. \cite{B}-\cite{C} described above, we shall
consider in this note the case in which the parity odd part of the
effective action is calculated for fermions on ${\mathcal S}^2\times
{\mathcal S}^1$ \,.
\subsection{Going from ${\mathcal S}^2 \times {\mathcal S}^1$ to ${\mathcal
S}^2$}
For the magnetostatic configurations, we see that the only $\tau$
dependence of the Dirac operator comes from $A_3$. This
dependence can, however, be eliminated by a redefinition of
the integrated fermion  fields. The set of allowed gauge
transformations in the imaginary time formalism is defined
in the usual way:
\bea
\psi(\tau,x) &\rightarrow& \exp[-i\Omega(\tau,x)] \psi(\tau,x),
\nonumber\\
\bar \psi(\tau,x) &\rightarrow& \exp[i\Omega(\tau,x)] \bar \psi(\tau,x),
\nonumber\\
A_\mu(\tau,x)  &\rightarrow& A_\mu(\tau,x) + \partial_\mu \Omega(\tau,x)
\eea
 where $\Omega(\tau,x)$ is a differentiable function vanishing at spatial
 infinity \mbox{$|x| \to \infty$}, and whose time boundary conditions
are chosen in order not to affect the fields' boundary
conditions. It turns out that $\Omega(\tau,x)$ can wind an arbitrary
number of times around the cyclic time dimension,
\be
\Omega(\tau,x) = \Omega(\tau,x) + {2\pi n}
\label{veintiunos}
\ee
where $n$ is an integer which labels the homotopy class of
the gauge transformation.

$Z[A]$ must be gauge invariant so that  we can compute it for time
independent gauge fields since one can always perform a gauge
transformation $A_\mu' = A_\mu + \partial_\mu\Lambda$ so that the $A'_\mu$
is time-independent. For the particular set of configurations
(\ref{eq:gauge}) such a transformation renders $A_3'$ constant. We see that
there is a family of $\Omega's$ achieving this while respecting the
boundary conditions \eqref{veintiunos},
\be
\Omega(\tau) = -\int_0^\tau d\tilde\tau A_3(\tilde\tau) +\frac1\beta\left(\int_0^\beta d\tilde\tau A_3(\tilde\tau) + 2\pi n
\right)\tau \;.
\ee
The freedom to choose $n$
could be used to further restrict the values of the constant
$A_3'$
 to a finite interval. In this sense, the value of the constant
in such an interval is the only ``essential'' i.e., gauge
invariant, $A_3$-dependent information contained in the configurations
(\ref{eq:gauge}), describing the gauge connection holonomy
$\exp(i \int_0^\beta d\tilde\tau A_3(\tilde\tau))$.

Let us, for the time being, disregard the case of large gauge
transformations (i.e., we restrict to the case $n=0$ in \eqref{veintiunos})
in order to avoid any assumption about large gauge invariance of the
fermionic measure in (\ref{eq:defgamma}) and then discuss this issue  on the
final results. Thus, the constant field $A_3'$  simply takes the mean value
of $\ A_3(\tau)$, $\tilde A_3 = \frac1\beta \int_0^\beta d\tilde\tau
A_3(\tau)$. Concerning the spatial components $A_i$, they remain $\tau$
independent after this transformation.

At this point we proceed to perform a Fourier transformation in the fermion
variables,
\bea
\psi(\tau,x) &=& \frac1\beta \sum_{n = -\infty}^{n = \infty} e^{i\omega_n\tau}\psi_n(x)\,, \nonumber\\
\bar\psi(\tau,x) &=& \frac1\beta \sum_{n = -\infty}^{n = \infty} e^{-i\omega_n\tau }\psi_n(x)\,,
\eea
where $\omega_n = (2n+1)\pi/\beta$ is the usual Matsubara frequency for the
case of fermion fields.  With this the  Dirac action  takes the form of an
infinite series of decoupled two-dimensional Euclidean Dirac   actions
\be
S_D = \frac1\beta \sum_{n = -\infty}^{n = \infty}\int_{S^2} d^2x \sqrt{g}
\,\bar\psi_n(x)( \not\!\!D_2 + M + i\gamma_3(\omega_n +  \tilde A_3))\psi_n(x)\;.
\label{eucli}
\ee
Here, $\not\!\!D_2$ denotes the 2-dimensional Dirac operator  acting on fermions
living in the sphere ${\mathcal S}^2$; it is given by
\be
\not\!\!D_2 = \gamma_a e^k_a(\partial_k +\frac14 \gamma_c \gamma_d \omega_{kcd} +i A_k(x))
\ee
Parametrizing ${\mathcal S}^2$ in terms of angular coordinates
$(x^1,x^2)=(\theta,\varphi)$, the induced metric becomes  ${\rm diag}\,g_{ij} =
 (R^2, R^2 \sin^2\theta)$, while the vierbeins $e_i^a$ ($a,b =1,2$) adopt
 the form:
 \be
 \begin{array}{lll}
 e^1_\theta = R\cos \varphi, & ~  & e^1_\varphi= -R \sin\varphi\sin\theta\\
 ~
 \\
e^2_\theta= R \sin\varphi, & ~ & e^2_\varphi= R \cos\varphi \sin\theta \;.
\end{array}
\ee
The only non-trivial component of the spin connection is
$\omega_{\varphi12} = (1-\cos\theta)$ so that
\be
\not\!\!D_2 = \gamma_a e^k_a \big(\partial_k + \frac14 i \delta^\varphi_k
\gamma_3 (1-\cos\theta) + i A_k(x) \big) \;.
\label{D2}
\ee
Thus, the action $S_D$ in \eqref{eucli} can be written in the form
\be
S_D = \frac1\beta \sum_{n = -\infty}^{n = \infty}\int_{S^2} d^2x \sqrt{g}
\,\bar \psi_n(x)\left(\vphantom{e^{a^n}}\!\not\!\!D_2 + \rho_n
\exp(i\gamma_3\phi_n)\right)\psi_n(x)\;,
\label{eucli2}
\ee
where
\be
\rho_n = (M^2 + (\omega_n +  \tilde A_3)^2)^{1/2} \; , \;\;\; \phi_n = \arctan\left(
\frac{\omega_n + \tilde A_3}{M}
\right) \;.
 \ee

Again, the path-integral measure factorizes:
\be
  \int D\bar\psi D\psi = \prod_{n = -\infty}^{n = \infty}  D\bar\psi_n(x) D\psi_n(x)
  \ee
 so that the partition function becomes a product
 \bea
 e^{-\Gamma[A,m]} &=& \prod_{n = -\infty}^{n = \infty} D\psi_n D\psi_n
 e^{-\int_{\mathcal S_2}d^2x \sqrt{g} \bar \psi_n(x)(\not\! D_2 +
	 \rho_n e^{i\gamma_3 \phi_n})\psi_n(x)} \nonumber\\
 &\equiv& \prod_{n = -\infty}^{n = \infty} \det (\not\!\!D_2 + \rho_n
 e^{i\gamma_3 \phi_n}) \;.
\label{f22}
 \eea
Now one can make in each pair of integrals in the infinite product the change of variables,
\be
\psi_n(x) = \exp[(-i\phi_n/2) \gamma_3] \psi_n'(x) \,, \;\;\;\; \bar \psi_n(x) = \bar\psi_n'(x) \exp[(-i\phi_n/2) \gamma_3] \,,
\label{f2222}
\ee
which can be seen as a a two-dimensional ``chiral rotation'' with
$\gamma_3$ identified as  the $d=2$ $\gamma_5$ matrix. Such changes
eliminates the phases $\phi_n$ in the Dirac operators at the cost of a
nontrivial  Fujikawa Jacobians $J_n$ associated to the chiral anomaly.
Indeed, each  Dirac operator determinant in \eqref{f22} is a product of
eigenvalues that grow with no bound thus requiring a regularization. One
can use for example  the heat-kernel regularization which consists in
introducing the identity
\be
I = \lim_{\Lambda \to \infty} \exp(-(\not\!\!D_2)^2/\Lambda^2)
\ee
using as regulating operator the one in the action that ensures gauge
invariance. One can easily see that the presence of the spin connection
contribution does not change the result of the Jacobian with respect to the
flat space case so that one finds
\be
J_n = \exp\left(-i\frac{e\phi_n}{2\pi} \int_{\mathcal S^2} d^2x  \varepsilon_{jk} \partial_j A_k\right)
\ee
so that finally one ends up with
\be
\exp(-\Gamma[A,m]) = \prod_{n = -\infty}^{n = \infty}
J_n\det (\not\!\!D_2 + \rho_n) \;.
\ee
Since the determinants in this formula do not depend on $m$ and our method
to calculate $\Gamma_{\rm odd}[A,m]$ consists in substracting the  positive
and negative mass results, they do not contribute to the parity odd
effective action.
In contrast, $\phi_n$ evidently depends on the sign of the mass. We then have
\be
\Gamma_{\rm odd} =  - \sum_{-\infty}^\infty \, \log J_n =
\frac{i}{2\pi}\sum_{-\infty}^\infty \phi_n \int_{\mathcal S^2} d^2x
\varepsilon_{jk} \partial_j A_k \;.
\ee
Performing the summation of the series, we obtain
\be
\Gamma_{\rm odd} = \frac{i}{2\pi} \frac{m}{|m|} \arctan\left(
\tanh\left( \frac{\beta |m|}{2}\right) \tan \frac12\int_0^\beta d\tau A_3(\tau)
\right)
\times \int_{\mathcal S^2} d^2x  \varepsilon_{jk} \partial_j A_k \;,
\label{finali}
\ee
which is the main result of this note.

Note that, again, invariance of the full effective action under large gauge
transformations may be achieved, by taking into account the anomalous term
with the proper sign.  Equation (\ref{finali}) not only does have the
proper zero-temperature limit, but it also produces a $\theta$-vacua term
in the opposite, high-temperature limit, an object known to be present in
the massive Schwinger model~\cite{cole}.  Indeed, we see that, when $\beta
m$ tends to zero, the leading non-trivial term is
\be \label{resulinf}
\Gamma_{\rm odd} [A,m] \sim i  \theta \int_{\mathcal S^2} d^2x
\varepsilon_{jk} \partial_j A_k \;,
\ee
with the dimensionless angle $\theta \equiv \frac{\beta^2 m \tilde A_3}{8
\pi}$.
This behaviour is to be expected, since one should obtain a result
corresponding to massive fermions in $2$ dimensions.

\section{Discussion}\label{sec:disc}
We have computed the exact mass-dependent parity-odd effective
action for fermions coupled to a $U(1)$ gauge field for a ${\mathcal S}^2
\times {\mathcal S}^1$ manifold. Once the standard anomalous
$\Gamma^{(0)}[A]= \frac12S_{CS}[A]$ contribution is incorporated, gauge
invariance holds even when large gauge transformations are considered.
Interestingly enough, the results coincides with the one obtained in
\cite{B} since the inclusion of the spin connection in the Dirac operator
does not affect the result.

Concerning the extension to the case of non-Abelian gauge symmetries, the
issue of large gauge invariance in non-Abelian finite temperature effective
actions for the case of an ${\mathbb R}^2\times\mathcal{S}^1$ manifold has
been discussed in \cite{DDGGSS}, and the explicit calculation of
$\Gamma_{\rm odd}[A]$  for fermions in the fundamental representation of
$SU(N)$  presented in \cite{C} so that we expect that the extension of the
present analysis to the non-Abelian case will follow the same steps for the
$\mathcal{S}^2\times\mathcal{S}^1$ manifold and lead to similar results as
those presented here.

Note that the flux quantization condition, namely, that
$\frac{\Phi_B}{2\pi} \in {\mathbb Z}$ can be altered when one considers a
path integral involving the insertion of matter field operators with
half-integer spins, since now the usual argument leading to Dirac's
quantization condition has to be applied to a field which is double-valued,
i.e., changes sign under $2\pi$ rotations. Thus, one can expect the value
of $\frac{\Phi_B}{2\pi}$ to be a half-integer in some cases.  Of course, a
full consideration has to be made of the contributions of the gauge field
action itself, when that field is dynamical, since it can also put constraints on the
allowed values of the flux (for example, by requiring finite energies).
Even in a situation when the flux is a half integer times $2 \pi$, large
gauge invariance may be restored by the presence of the anomalous term with
the proper sign {\em and coefficient}. In the context of Pauli-Villars
regularization, that would correspond to having three, rather
than one, regulator fields. This number of regulators, is not required in the
calculation of the fermion loop (where one is sufficient); rather, it is
needed in order to regulate diagrams with operator insertions, where the
superficial degree of divergence is increased.

It is important to note that the reduction of the $2+1$ dimensional system to a
collection of $2$-dimensional ones, for the case of a static metric like the
one in (\ref{eq:met}), can certainly be generalized to spatial manifolds
different from the sphere. Indeed, the main properties we relied upon to
achieve the reduction still hold true, in particular, the fact that the
vierbeins are effectively two dimensional, they do not mix with the time-like
coordinate.

Let us conclude by pointing out that the knowledge of the fermionic
determinant is an essential ingredient in the path-integral approach to
bosonization in $d \geq 2$ dimensions, as developed in
refs.\cite{FS}-\cite{moreno}. Indeed, in $3$ space-time dimensions, the fermionic
action and its $\bar\psi \gamma^\mu \psi$ current become related, by a
duality transformations, to an effective low energy Chern-Simons action and
a conserved dual current on the bosonic side. Thus, we believe, the issue
of having non perturbative results for the fermionic determinant may be
relevant to the correct implementation of the bosonization/duality programs at a
finite temperature, within the context of condensed matter physics.
We have shown that the fermionic determinant leads to a non-local
gauge action which reduces to a CS action solely in the $T\to 0$
limit and this takes place both for the spatial coordinates taking values
in ${\mathbb R}^2$ or ${\mathcal S}^2$. Moreover, it becomes a $\theta$-vacua
term when the opposite, high temperature limit is considered.

Finally, note also that the program of extending the study of dualities to
different manifolds could also be considered in higher dimensions: indeed, in
$d>3$ dimensions, it has been shown that the dual current involves a
$d-2$ Kalb-Ramond field \cite{Fo}. Regarding the massless fermion case,
different dualities  connect the fermionic theory with vector and scalar
fields theories \cite{Moreno2} and one can even consider the case in which
fermions are coupled to a gravitational background \cite{MF}.
We hope to report on the result of that extension, as well as on the
consequences for condensed matter applications, in future works.

~

\noindent ACKNOWLEDGEMENTS

This work was supported by ANPCyT, CICBA, CONICET,
UBA, UNCuyo and UNLP.


\begin{thebibliography}{99}
\bibitem{Hasan} M.~Z.~Hasan and C.~L.~Kane,   Rev.\ Mod.\ Phys.\ {\bf 82}
	(2010) 3045.
\bibitem{Widual}  N.~Seiberg, T.~Senthil, C.~Wang and E.~Witten,
  %``A Duality Web in 2+1 Dimensions and Condensed Matter Physics,''
  Annals Phys.\  {\bf 374} (2016) 395.
\bibitem{Karch} A.~Karch and D.~Tong,
  %``Particle-Vortex Duality from 3d Bosonization,''
  Phys.\ Rev.\ X {\bf 6} (2016) no.3,  031043
\bibitem{Nastase} J.~Murugan and H.~Nastase,
  %``Particle-vortex duality in topological insulators and superconductors,''
  arXiv:1606.01912 [hep-th].
  %``Particle-vortex duality in topological insulators and superconductors,''
\bibitem{directions}  P.~S.~Hsin and N.~Seiberg,
  %``Level/rank Duality and Chern-Simons-Matter Theories,''
  JHEP {\bf 1609} (2016) 095;
  Ð.~Radicevic, D.~Tong and C.~Turner,
  %``Non-Abelian 3d Bosonization and Quantum Hall States,''
  JHEP {\bf 1612} (2016) 067; A.~Karch, B.~Robinson and D.~Tong,
  %``More Abelian Dualities in 2+1 Dimensions,''
  JHEP {\bf 1701} (2017) 017;  A.~Cappelli, E.~Randellini and J.~Sisti,
  %``Three-dimensional Topological Insulators and Bosonization,''
  arXiv:1612.05212 [cond-mat.str-el].
  %%%%%%%%%%%%%%%%%%%%%%%%%%%%%%%%%%%%%%%%%%%%%%%%%%%%%%
  \bibitem{GRS} R.~E.~Gamboa Sarav\'{\i}, G.~L.~Rossini, and F.~A.~Schaposnik, Int.\ J.\ Mod.\ Phys.\ A {\bf 11}, 1996 2643.
        \bibitem{Redlich} A.~N.~Redlich, Phys.\ Rev.\ Lett. {\bf 52} (1984) 18 ; Phys.\ Rev.\ D {\bf 29} (1984) 1984.
      \bibitem{Dunne}  G.~V.~Dunne,
  %``Aspects of Chern-Simons theory,''
  hep-th/9902115.
\bibitem{A} S.~Deser, L.~Griguolo and D.~Seminara,
  %``Gauge invariance, finite temperature and parity anomaly in D = 3,''
  Phys.\ Rev.\ Lett.\  {\bf 79} (1997) 1976
\bibitem{B} C.~Fosco, G.~L.~Rossini and F.~A.~Schaposnik,
  %``Induced parity breaking term at finite temperature,''
  Phys.\ Rev.\ Lett.\  {\bf 79} (1997) 1980.
\bibitem{C} C.~D.~Fosco, G.~L.~Rossini and F.~A.~Schaposnik,
  %``Abelian and nonAbelian induced parity breaking terms at finite temperature,''
  Phys.\ Rev.\ D {\bf 56} (1997) 6547.
        \bibitem{Dunne}  G.~V.~Dunne,   %``Aspects of Chern-Simons theory,''
  hep-th/9902115.
 \bibitem{cole} S.~R.~Coleman,
  %``More About the Massive Schwinger Model,''
  Annals Phys.\  {\bf 101}, 239 (1976).
  \bibitem{DDGGSS} S.~Deser, L.~Griguolo and D.~Seminara,
  %``Large gauge invariance in nonAbelian finite temperature effective actions,''
  Phys.\ Rev.\ D {\bf 67} (2003) 065016.
  \bibitem{FS} E.~H.~Fradkin and F.~A.~Schaposnik,
  %``The Fermion - boson mapping in three-dimensional quantum field theory,''
  Phys.\ Lett.\ B {\bf 338} (1994) 253
  \bibitem{moreno} J.~C.~Le Guillou, E.~Moreno, C.~Nunez and F.~A.~Schaposnik,
  %``NonAbelian bosonization in two-dimensions and three-dimensions,''
  Nucl.\ Phys.\ B {\bf 484} (1997) 682;
  %``On three-dimensional bosonization,''
  Phys.\ Lett.\ B {\bf 409} (1997) 257.
  \bibitem{Fo} C.~D.~Fosco and F.~A.~Schaposnik,
  %``Bosonization of vector and axial vector currents in (3+1)-dimensions,''
  Phys.\ Lett.\ B {\bf 391} (1997) 136.
  \bibitem{Moreno2}E.~F.~Moreno and F.~A.~Schaposnik,
  %``Dualities and bosonization of massless fermions in three dimensional space-time,''
  Phys.\ Rev.\ D {\bf 88} (2013) no.2,  025033
  \bibitem{MF}C.~D.~Fosco, E.~F.~Moreno and F.~A.~Schaposnik,
  %``Fermions in an AdS$_3$ Black Hole Background and the Gauge-Gravity Duality,''
  Phys.\ Rev.\ D {\bf 85} (2012) 046005.
 \end{thebibliography}
  \end{document}